\documentclass[fleqn,twoside]{article}
\usepackage{espcrc2}

\usepackage{graphicx}
\usepackage[figuresright]{rotating}

\RequirePackage{xspace}

\newcommand{\AmS}{{\protect\the\textfont2
  A\kern-.1667em\lower.5ex\hbox{M}\kern-.125emS}}
\newcommand{\Dbar}{\kern 0.18em\overline{\kern -0.18em D}{}\xspace}

\newcommand{\Dz}{\ensuremath{D^0}\xspace}
\newcommand{\Dzb}{\ensuremath{\Dbar^0}\xspace}
\newcommand{\DzDzb}{\ensuremath{\Dz {\kern -0.16em \Dzb}}\xspace}
\newcommand{\DsP}{\ensuremath{D_S^+}\xspace}
\newcommand{\DsM}{\ensuremath{D_S^-}\xspace}
\newcommand{\DspDsm}{\ensuremath{\DsP {\kern -0.16em \DsM}}\xspace}
\newcommand{\Dp}{\ensuremath{D^+}\xspace}
\newcommand{\Dm}{\ensuremath{D^-}\xspace}
\newcommand{\DpDm}{\ensuremath{\Dp {\kern -0.16em \Dm}}\xspace}
\newcommand{\psp}{\psi^{\prime}}

\newcommand{\jpsi}{J/\psi}

\newcommand{\chicz}{\chi_{c0}}

\newcommand{\ppkk}{\pi^+\pi^-K^+K^-}
\newcommand{\EE}{e^+e^-}

\newcommand{\pip}{\pi^+}
\newcommand{\pim}{\pi^-}
\newcommand{\piz}{\pi^0}

\newcommand{\kap}{K^+}
\newcommand{\kam}{K^-}
\newcommand{\ks}{K^0_s}

\title{Study of Scalar Mesons at BES-II}
\author{Haibo Li\address{Institute of High Energy Physics,
           P.O.Box 918, Beijing  100049, China} \thanks{Email:
lihb@ihep.ac.cn}}
\begin{document}
\begin{abstract}
Recent results from BES-II experiment on hadron
spectroscopy using $\jpsi$ and $\psp$ data samples collected in $\EE$
annihilation are presented,
including study of the scalar particles in $\jpsi$ radiative
and hadronic decays, 
the observation of $X(1810)$ in $J/\psi \to \gamma \phi \omega$, 
as well as pair productions of scalars in $\chicz$ hadronic decays.
\vspace{1pc}
\end{abstract}

\maketitle

\section{Introduction}

The scalar mesons are one of the most controversial subjects in hadron
physics.  Below 1.0 GeV, there are two $I = 0$ saclar candidates, $\sigma$
($\pi\pi$ S-Wave), $f_0(980)$ and one $I = 1/2$ $K\pi$ S-wave, $\kappa$,
in the Particle Data Group (PDG) lists~\cite{pdg}. Between 1.0 GeV and 2.2 
GeV, PDG lists the following $I = 0$ scalar states: $f_0(1370)$, $f_0(1500)$
$f_0(1710)$ and one $I = 1/2$ scalar state: $K^*_0(1430)$. Scalar mesons
have been traditionally studied in scattering experiments. However, in these 
experiments the mesons can be difficult to disentangle from non-resonant
backgrounds. Radiative and hadronic $J/\psi$ decays provide an excellent
laboratory to probe these states. 

Recently, based on 58 million $J/\psi$ decay events and 14 million
$\psi(2S)$ decay events collected with BES-II detector~\cite{bes-ii},  
many detailed partial wave analyses (PWA) have been performed in order
 to understand the structure, decays and production of the scalar mesons. 
In this paper, we present some of the results from such study at BES-II. 

\section{Scalars in $J/\psi$ Radiative and Hadronic Decays}

\begin{figure}[htbp]
\centering{
\includegraphics[width=7.0cm,height=12.0cm]{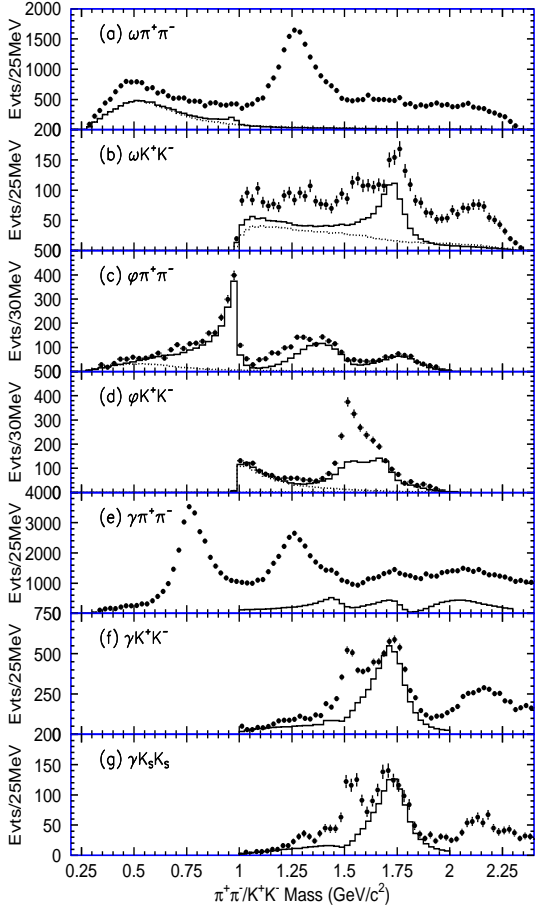}
}
\caption{The invariant mass distributions of the pseudoscalar meson
pairs recoiling against $\omega$, $\phi$, or $\gamma$ in $J/\psi$
decays measured at BES-II. The dots with error bars are data, the
solid histograms are the scalar contribution from PWA, and the
dashed lines in (a) through (c) are contributions of $\sigma$ from
the fits, while the dashed line in (d) is the $f_0(980)$. Notice
that not the full mass spectra are analyzed in (e), (f), and (g).
Results in (e) are preliminary, otherwise are published}
\label{scalars}
\end{figure}

Using the world largest $\jpsi$ data sample
in $e^+e^-$ annihilation
experiment, BES-II studied the scalars decay into pair of
pseudoscalars ($\pip\pim$, $\piz\piz$, $\kap\kam$ and $\ks\ks$) in
$\jpsi$ radiative decays as well as recoiling against a $\phi$ or
an $\omega$~\cite{wpipi,wkk,phipipi,gkk}. The full mass spectra
and the scalar part in them are shown in Fig.~\ref{scalars}.

\subsection{$\sigma$ and $\kappa$}

From the analyses, BES-II sees significant contributions of $\sigma$
particle in $\omega\pip\pim$ and $\omega\kap\kam$, and also hint
in $\phi\pip\pim$. 
In $J/\psi \rightarrow \omega \pi^+\pi^-$, there are conspicuous $\omega
f_2(1270)$ and $b_1(1235)\pi$ signals. At low $\pi\pi$ mass, a
large, broad peak due to the $\sigma$ is observed as shown in
Fig.~\ref{scalars}(a). 
Two independent partial wave analyses are
performed on $\omega\pip\pim$ data and four different
parameterizations of the $\sigma$ amplitude are tried~\cite{wpipi}, all give
consistent results for the $\sigma$ pole, which is at  $(541 \pm
39) - i(252 \pm 42)$~MeV/c$^2$.  
Recently, an analysis of $\psi(2S) \rightarrow \pi^+\pi^- J/\psi$ has been
performed to study the $\sigma$~\cite{zhuys}. The pole position of
$\sigma$ is consistent with that from $J/\psi \rightarrow \omega
\pi^+\pi^-$. 

Events over all of the
4-body phase space for $J/\psi \rightarrow K^+ K^- \pi^+\pi^-$ have been
fitted. We find evidence for the $\kappa$ in the process $J/\psi \rightarrow 
 \overline{K}^*(892)^0 \kappa$, $\kappa \rightarrow (K\pi)_S$.  We
select a $K^- \pi^+$ pair in the $\overline{K}^*(892)^0$ mass range
$892 \pm 100$ MeV.  Two independent PWA,
by the covariant helicity amplitude method~\cite{34} and by the variant
mass and width method~\cite{24}  have been
performed, providing a cross check with each other. They reproduce the
data well, and the results are in good agreement. The low mass enhancement
is well described by the scalar $\kappa$, which is highly required
in the analyses. Parameter values of Breit-Wigner (BW) mass and width
of the $\kappa$, averaged from those obtained by the two
methods, are
$878\pm23^{+64}_{-55}$ MeV/$c^2$ and $499\pm 52^{+55}_{-87}$
MeV/$c^2$. The pole position is determined to be
$(841\pm 30^{+81}_{-73})-i(309 \pm
45^{+48}_{-72})$ MeV/$c^2$~\cite{kappa_paper}. 

\subsection{$f_0(980)$ in $J/\psi$ Decays}

Strong $f_0(980)$ is seen in $J/\psi \rightarrow \phi \pi ^+\pi ^-$ and $\phi
K^+K^-$ modes~\cite{phipipi}, from which the resonance parameters are measured to be $M =
965 \pm 8(stat) \pm 6(syst) $~MeV/c$^2$, $g_1 = 165 \pm 10(stat)
\pm 15(syst)$~MeV/c$^2$ and $g_2/g_1 = 4.21 \pm 0.25(stat) \pm
0.21(syst)$, where $M$ is the mass, and $g_1$ and $g_2$ are the
couplings to $\pi\pi$ and $K\bar{K}$ respectively if the
$f_0(980)$ is parameterized using the the Flatt\'e's formula. The
production of $f_0(980)$ is very weak recoiling against an
$\omega$ or a photon, which indicates $s\bar{s}$ is the dominant
component in it.
 
\subsection{Scalar above $1.0$ GeV in $J/\psi$ Decays}

 In $J/\psi \rightarrow \phi \pip \pim$ decay, a scalar contribution near
1.4 GeV on $\pi^+\pi^-$ invariant mass distribution is found as shown in
Fig.~ref{scalars}(c), it is due to  the
dominant $f_0(1370)$
interfering with a smaller $f_0(1500)$ component. The mass and
width of $f_0(1370)$ are determined to be: $M = 1350 \pm
50$~MeV/c$^2$ and $\Gamma = 265 \pm 40$~MeV/c$^2$. 
In $\gamma \pip
\pim$, a similar structure is observed in the same mass region,
the fit yields a resonance at mass $1466\pm 6(stat)\pm
16(syst)$~MeV/c$^2$ with width of $108^{+14}_{-11}(stat) \pm
21(syst)$~MeV/c$^2$, possibly the $f_0(1500)$~\cite{phipipi}, and the
contribution from the $f_0(1370)$ can not be excluded. While, the
production of $f_0(1370)$ and $f_0(1500)$ in $\gamma K \bar{K}$ is
insignificant~\cite{gkk}. 

The $K^+K^-$ invariant mass distributions from $\gamma K \bar{K}$
and $\omega K^+K^-$, the $\pip\pim$ invariant mass distributions
from $\gamma \pip\pim$, and $\phi\pip\pim$ show clear scalar
contribution around 1.75~GeV/c$^2$. Two states are resolved from
the bump, one is $f_0(1710)$ with $M \sim 1740$~MeV/c$^2$ and
$\Gamma\sim 150$~MeV/c$^2$ which decays to $K \bar{K}$ mostly, and
one possible new state $f_0(1790)$ with $M \sim 1790$~MeV/c$^2$
and $\Gamma\sim 270$~MeV/c$^2$ which couples to $\pi\pi$ stronger
than to $K\bar{K}$. However, the existence of the second scalar
particle needs confirmation: the signal observed in $\phi
f_0(1790)$ is rather in the edge of the phase space, and the
reconstruction efficiency of the $\phi$ decreases dramatically as
the momentum of the $\phi$ decreases thus the momentum of the kaon
from $\phi$ decays is very low and can not be detected~\cite{zhuys}.
Furthermore, there are wide higher mass scalar states above
2~GeV/c$^2$ as observed in $\gamma \pip\pim$ (Fig.~\ref{scalars}e)
and $\gamma K \bar{K}$~\cite{pdg}, whose tails may interfere with
the $f_0(1710)$ and produce structure near the edge of the phase
space.

The discussions of these measurements for understanding the nature of the
scalar particles can be found in
Refs.~\cite{oset,close,zhao}, where the $\jpsi$ decay
dynamics and the fractions of the possible $q\bar{q}$ and glueball
components in the states are examined.

\section{Observation of Threshold Enhancements in $J/\psi$ Decays}

In the last few years, anomalous enhancements near threshold
in the invariant
mass spectra of $p \bar p$ and $p \bar \Lambda$ pairs were observed 
in $J/\psi \to \gamma p \bar p$ \cite{gppb} and $J/\psi \to p K \bar
\Lambda$ \cite{pkl} decays, respectively, by the BES-II experiment.
These surprising experimental observations stimulated many theoretical
speculations. Therefore it is of special interests to search for
possible resonances in other baryon-antibaryon, baryon-meson, and
meson-meson final states.
\begin{figure}[htb]
\includegraphics[width=6.0cm,height=6.0cm]{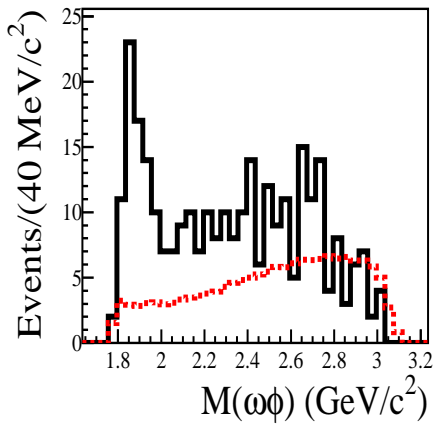}
\caption{The $K^+K^-\pi^+\pi^-\pi^0$ invariant mass distribution for the
$J/\psi\to \gamma \omega\phi$ candidate events. The dashed curve indicates
the acceptance varying with the $\omega\phi$ invariant mass.}
\label{g-phi-omega}
\end{figure}
\begin{figure}[htb]
\includegraphics[width=6.0cm,height=6.0cm]{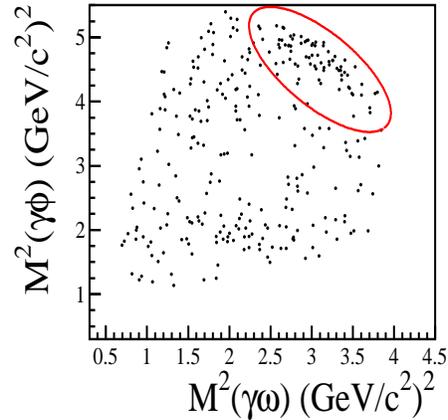}
\caption{$M^2(\gamma \phi)$ .vs. $M^2(\gamma \omega)$, Dalitz plot in $J/\psi \rightarrow \gamma \phi \omega$ decay.}
\label{g-phi-omega-2}
\end{figure}

\subsection{$X(1810)$ in $J/\psi \rightarrow \gamma \phi \omega$ Decays}

An enhancement near threshold is observed in the $\omega \phi$
    invariant mass spectrum from the doubly OZI suppressed decays of 
    $J/\psi \to \gamma \omega \phi$, where $\omega$ and $\phi$ are
reconstructed from $\pi^+\pi^-\pi^0$ and
$K^+K^-$ final
states~\cite{gphiomega}, respectively. Figure~\ref{g-phi-omega} shows   
the $K^+K^-\pi^+\pi^-\pi^0$ invariant
mass distribution for events with $K^+K^-$ invariant mass within the nominal 
$\phi$ mass range ($|m_{K^+K^-}-m_{\phi}|<15$ MeV/$c^2$) and the  
$\pi^+\pi^-\pi^0$ mass within the $\omega$ mass range ($|m_{\pi^+\pi^-\pi^0}
-m_{\omega}|<30$ MeV/$c^2$), and a structure peaked near  $\omega\phi$ 
threshold is observed.   The dashed curve in the figure indicates
how the acceptance varies with invariant mass. The peak is also evident
as a diagonal band along the upper right-hand 
edge of the Dalitz plot in Fig.~\ref{g-phi-omega-2}. There is also a horizontal band 
near  $m^2_{\gamma K^+ K^-} = 2$ (GeV/$c^2$)$^2$ in the Dalitz  plot, which 
mainly comes from background due to $J/\psi \to \omega K^*K$. 
\begin{figure}[hpb]
\includegraphics[width=6.0cm,height=6.0cm]{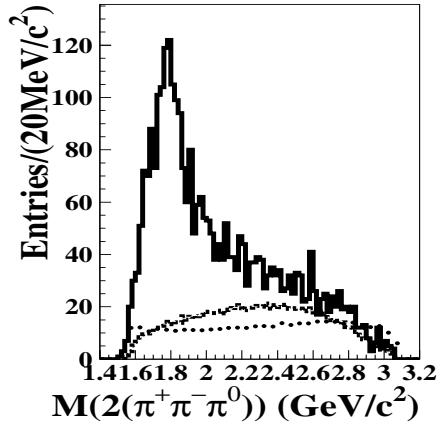}
\caption{The $2(\pi^+ \pi^- \pi^0)$
invariant mass distribution for candidate events. The dashed
curve is the phase space invariant mass distribution, and the dotted curve
shows the acceptance versus the $\omega\omega$
invariant mass.}
\label{fig2}
\end{figure}
\begin{figure}[hpb]
\includegraphics[width=6.0cm,height=5.0cm]{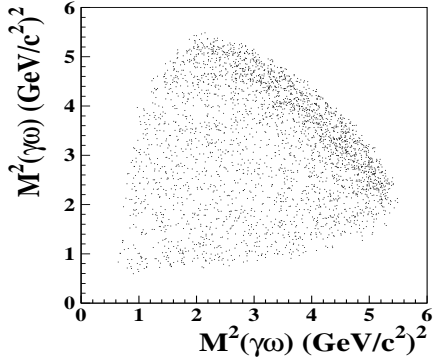}
\caption{The Dalitz plot in $J/\psi \rightarrow \gamma \omega \omega$ decays. }
\label{fig2-2}
\end{figure}

From a PWA with covariant helicity 
coupling amplitudes, the spin-parity of the $X = 0^{++}$ with an 
$\mathcal{S}$-wave $\omega\phi$ system is favored. 
The enhancement near is observed with a statistical significance of more 
than 10$\sigma$.  The mass and width of 
the enhancement are determined to be $M = 1812^{+19}_{-26}$ (stat) $\pm$ 
18 (syst) MeV/$c^2$ and $\Gamma = 105 \pm 20$ (stat) $\pm$ 28 (syst)
MeV/$c^2$, and the product branching fraction is $\mathcal{B}(J/\psi\to\gamma X)\cdot
\mathcal{B}(X\to\omega\phi)$ = (2.61 $\pm$ 0.27 (stat) $\pm$  0.65 (syst)) 
$\times$ $10^{-4}$.  The mass and width of this state are not
compatible with 
any known scalars listed in the PDG~\cite{pdg}. It 
could be an unconventional state \cite{liba,lixq,pedro,chaokt,bugg}. 
However, more statistics and  further 
studies are needed to clarify this.
\begin{figure}[htb]
\centering{
\includegraphics[width=5.6cm,height=5.5cm]{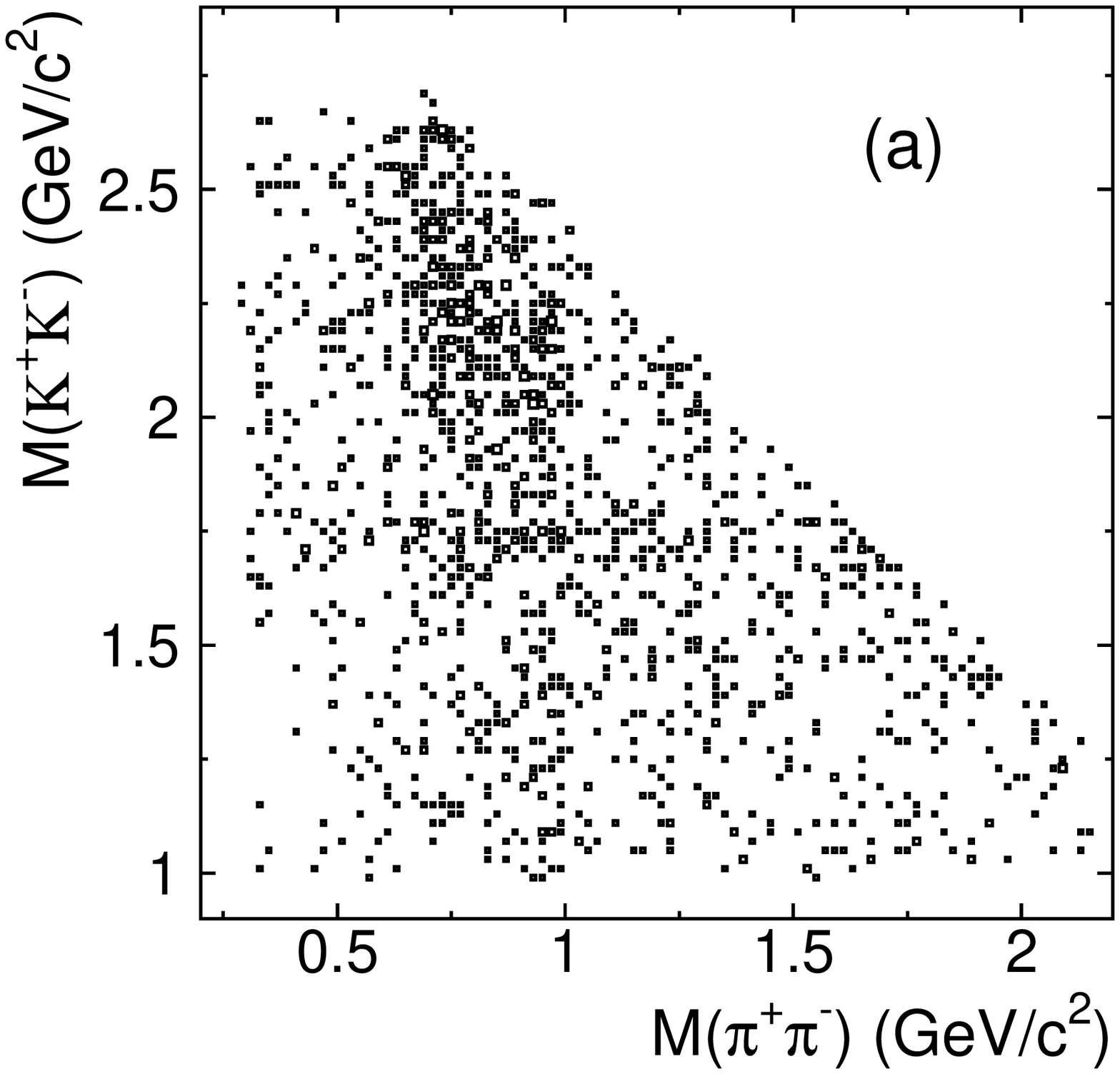}
\includegraphics[width=5.6cm,height=5.5cm]{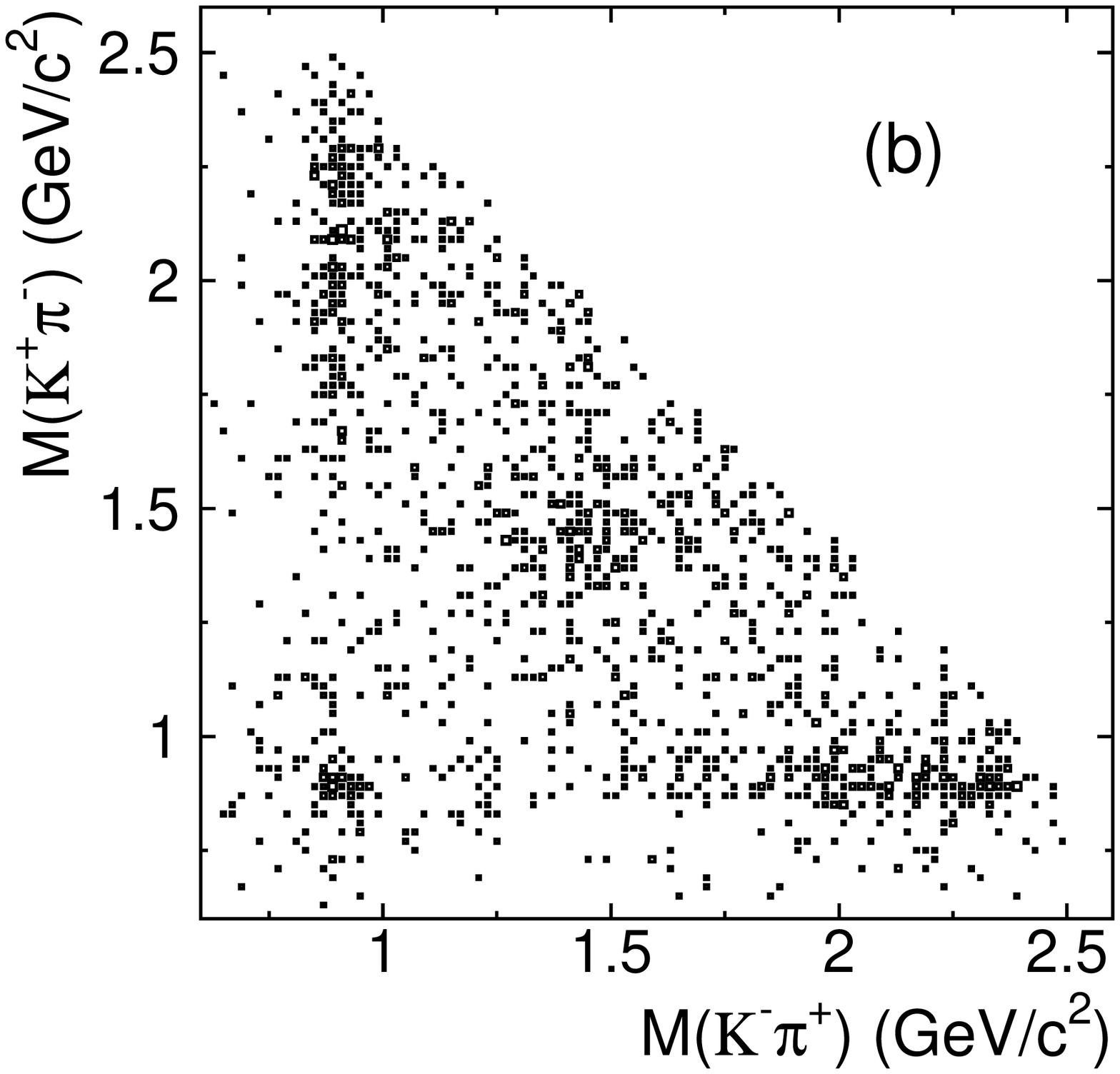}
}
\caption{The scatter plots of (a) $K^+K^-$
versus $\pi^+\pi^-$ and (b) $K^+\pi^-$ versus $K^-\pi^+$ invariant
mass for selected $\psi^{\prime} \rightarrow \gamma\chi_{c0}$,
$\chi_{c0} \rightarrow \pi^+\pi^-K^+K^-$ events.}
\label{fig_xchi0}
\end{figure}

\subsection{Enhancement in $J/\psi \rightarrow \gamma \omega \omega$ Decays}

The decay mode $J/\psi \rightarrow  \gamma \omega \omega$, $\omega
\rightarrow \pi^+\pi^-\pi^0$ is analyzed using a sample of 5.8 million
$J/\psi$ decay events. The $\omega \omega$ invariant mass distribution peaks
 at 1.76 GeV/$c^2$, just above the $\omega \omega$
threshold. The histogram of Fig. \ref{fig2} shows the  $2(\pi^+ \pi^-
\pi^0)$
invariant mass distribution of events with both $
\pi^+ \pi^- \pi^0$ masses within the $\omega$ range ($|m_{
\pi^+\pi^-\pi^0 }-m_{\omega}| <$40 MeV/$c^2$). There are 3046
events with a
clear peak at 1.76 GeV/$c^2$. The phase space invariant mass
distribution 
and the acceptance versus $\omega \omega$ invariant mass
are also shown in the figure. The corresponding Dalitz plot is
shown in Fig. \ref{fig2-2}.
\begin{table*}[htb]
\caption {Summary of the $\chicz\to \ppkk$ results, where $X$
represents the intermediate decay modes, $N^{obs}$ is the number
of fitted events, and s.s. indicates signal significance.}
\label{pwachi}
\begin{tabular}{@{}lllll}\hline

Decay Mode  &  $N^{obs}$   & ${\cal BR}$($10^{-4}$)  &  s.s     \\
(X)      &                 & ${\cal BR}(\chi_{c0} \rightarrow X \rightarrow \pi^+\pi^- K^+K^-)$ &  \\
\hline

${f_0(980)f_0(980)}$&$27.9\pm8.7$&$3.46\pm
1.08^{+1.93}_{-1.57}$&$5.3\sigma$\\
${f_0(980)f_0(2200)}$&$77.1\pm13.0$&$8.42\pm1.42^{+1.65}_{-2.29}$&$7.1\sigma$
\\
${f_0(1370)f_0(1710)}$&$60.6\pm15.7$&$7.12\pm1.85^{+3.28}_{-1.68}$&$6.5\sigma$\\
${K^*(892)^0\bar
K^*(892)^0}$&$64.5\pm13.5$&$8.09\pm1.69^{+2.29}_{-1.99}$&$7.1\sigma$\\
${K^*_0(1430)\bar
K^*_0(1430)}$&$82.9\pm18.8$&$10.44\pm2.37^{+3.05}_{-1.90}$&$7.2\sigma$\\
${K^*_0(1430)\bar K^*_2(1430)} +
c.c.$&$62.0\pm12.1$&$8.49\pm1.66^{+1.32}_{-1.99}$&$8.7\sigma$\\
${K_1(1270)^{+}K^{-} + c.c.,}$&&&\\
~~~$K_1(1270)\to
K\rho(770)$&$68.3\pm13.4$&$9.32\pm1.83^{+1.81}_{-1.64}$&$8.6\sigma$\\
${K_1(1400)^{+}K^{-} + c.c.,}$&&&\\
~~~$K_1(1400)\to
K^*(892)\pi$&$19.7\pm8.9$&$< 11.9$ (90\% C.L.) &$2.7\sigma$\\
\hline
\end {tabular}
\end {table*}

The partial wave
analysis shows a strong contribution from $0^-$ for the $\omega\omega$
invariant mass below 2.0 GeV/$c^2$, and with 
with small contributions from $f_0(1710)$, $f_2(1640)$, and
$f_2(1910)$.. Therefore the study of the
$\eta(1760)$ is the main goal of this analysis. 
The mass of the pseudoscalar is $M$ = 1744 $\pm$ 10 (stat)
$\pm$ 15 (syst) MeV/$c^2$, the width $\Gamma$ =
$244^{+24}_{-21}$ (stat) $\pm$ 25 (syst) MeV/$c^2$, and the product
branching fraction is Br($J/\psi\to\gamma\eta(1760)$) $\cdot$
Br($\eta(1760)\to \omega\omega$) = (1.98 $\pm$ 0.08 (stat)  $\pm$ 0.32
(syst))
$\times$ 10$^{-3}$. 

\section{Pair Productions of Scalars in $\chi_{c0} \rightarrow
\pi^+\pi^- K^+ K^-$}

PWA of $\chi_{c0} \rightarrow \pi^+ \pi^- K^+
K^-$ is performed~\cite{ppkk} using $\chicz$ produced in $\psp$
decays at BES-II. In 14~M produced $\psp$ events, 1371
$\psi^{\prime} \rightarrow \gamma \chi_{c0}$, $\chi_{c0}
\rightarrow \pi^+\pi^-K^+K^-$ candidates are selected with around
3\% background contamination.

Figure~\ref{fig_xchi0}(a) shows the scatter plot of $K^+K^-$ versus
$\pi^+\pi^-$ invariant mass which provides further information on
the intermediate resonant decay modes for $(\pi^+\pi^-)(K^+K^-)$
decay, while Fig.~\ref{fig_xchi0}(b) shows the scatter plot of
$K^+\pi^-$ versus $K^-\pi^+$ invariant masses for the resonances
with strange quark.

Besides $(\pi\pi)(KK)$ and $(K\pi)(K\pi)$ modes, $(K\pi\pi)K$ mode
which leads to a measurement of $K_1(1270)K$ and $K_1(1400)K$
decay processes is also included in the fit. The PWA results are
summarized in Table~\ref{pwachi}. From these results, we notice
that scalar resonances have larger decay fractions compared to
those of tensors, and such decays provide a relatively clean
laboratory to study the properties of scalars, such as $f_0(980)$,
$f_0(1370)$, $f_0(1710)$, and so on. 

The above results supply important information on the
understanding of the natures of the scalar states~\cite{zhaoq}.

\end{document}